\documentclass[twocolumn,showpacs,preprintnumbers,amsmath,amssymb]{revtex4}

\usepackage{graphicx}
\usepackage{dcolumn}
\usepackage{bm}
\newcommand{\stf}[1]{{\langle {#1} \rangle}} 

\begin{document}


\title{Carter-like constants of the motion in Newtonian gravity and electrodynamics}

\author{Clifford M. Will}
\email{cmw@wuphys.wustl.edu}
\homepage{http://physics.wustl.edu/cmw} 
\affiliation{%
GReCO, Institut d'Astrophysique de Paris, CNRS,\\
Universit\'e Pierre et Marie Curie, 98 bis Bd. Arago, 75014 Paris, France \\
Department of Physics, McDonnell Center for the Space Sciences \\
Washington University, St. Louis MO 63130 USA\footnote{Permanent address}
}%

\date{\today}

\begin{abstract}
For a test body orbiting an axisymmetric body in Newtonian gravitational theory with multipole moments $Q_\ell$, (and for a charge in a non-relativistic orbit about a charge distribution with the same multipole moments) we show that there exists, in addition to the energy and angular momentum component along the symmetry axis, a conserved quantity analogous to the Carter constant of Kerr spacetimes in general relativity, if the odd-$\ell$ moments vanish, and the even-$\ell$ moments satisfy $Q_{2\ell} = m (Q_2/m)^\ell$.  Strangely, this is precisely the relation among mass moments enforced by the no-hair theorems of rotating black holes.  
By contrast, if Newtonian gravity is supplemented by a multipolar gravitomagnetic field, whose leading term represents frame-dragging (or if the electrostatic field is supplemented by a multipolar magnetic field), we are unable to find an analogous Carter-like constant.  This further highlights the very special nature of the Kerr geometry of general relativity.

\end{abstract}

\maketitle

\section{\label{sec:level1}Introduction}

The Kerr geometry of the rotating black hole in general relativity has long been notable for its remarkable properties (see, for example \cite{chandra,mtw}).  Apart from being the unique, stationary, axisymmetric, asymptotically flat vacuum solution of Einstein's equations with a non-singular event horizon~\cite{carter72,robinson}, it manifests additional deep symmetries that are still not fully understood.  One expression of this is the statement that the Kerr black hole has ``no hair''; its exterior spacetime geometry is completely determined by two parameters, mass $m$ and spin parameter $a$ (the angular momentum per unit mass).   For example, it is possible to characterize the exterior spacetime of Kerr by a set of electric, or mass multipoles, $Q_\ell$, and magnetic, or current multipoles, $J_\ell$, where $\ell \ge 0$ is an angular harmonic index.  The near baldness of the geometry is expressed efficiently by the equation~\cite{hansen}
\begin{equation}
Q_\ell + {\rm i} J_\ell = m({\rm i}a)^\ell \,,
\label{nohair}
\end{equation}
with $Q_0$ being the mass $m$, and $J_1 = S = ma$ being the angular momentum, in units where $G=c=1$.  As a consequence, for example $Q_2 = -ma^2$, and $Q_{2\ell} = m(Q_2/m)^\ell$.   

Another remarkable property of the Kerr geometry is the existence of an additional constant of the motion for the geodesic of a test body, beyond the trivially conserved rest mass $\mu$ of the particle, the energy $E$, and the component of angular momentum $L_z$ along the black hole's symmetry axis.  The conservation of $E$ and $L_z$ arise simply from the stationarity and axisymmetry of the geometry.  The additional constant of the motion, called the ``Carter'' 
constant~\cite{carter}, is unexpected, since the geometry is not spherically symmetric.  It is a combination of the total orbital angular momentum $L^2$, which itself is not conserved for generically oriented orbits, components of the particle's linear momentum (squared), and other variables dependent upon the angular momentum parameter $a$ and the angle $\theta$ between the particle's position and the symmetry axis~\cite{defelice,rosquist}.  The existence of this added constant is intimately linked to the precise relationship among the moments that characterize the deviations from spherical symmetry, expressed by Eq.\ (\ref{nohair}).  Mathematically, it owes its constancy to the existence of a so-called ``Killing tensor'' $\xi_{\alpha\beta}$, a generalization of the Killing vectors $\xi_\alpha$ that generate the time and azimuthal symmetries that lead to the conservation of $E$ and $L_z$~\cite{walker}.  

A consequence of the existence of the Carter constant is that it makes  the geodesic equation in Kerr completely integrable, that is, solvable in terms of quadratures.  In practical terms, this may be a useful tool in analyses of the orbits of small black holes around massive black holes.  These are the extreme mass-ratio inspirals (EMRI) that are an important potential source of gravitational waves for the proposed space interferometer {\em LISA}.  Considerable work has gone into trying to determine how the Carter constant evolves with time because of gravitational radiation reaction, and how that plays into the evolution of inclined, inspiralling orbits around rotating black holes~\cite{mino,drasco05,hughes05,sago}.  

In a recent paper, Flanagan and Hinderer~\cite{flanhind} studied the evolution of the Carter constant using a post-Newtonian multipole approach for the effects of radiation reaction.  
In that paper, they pointed out the unusual fact that a Carter-like constant can exist even in Newtonian gravitational theory.  For an axisymmetric body characterized by a  mass $m$ and quadrupole moment $Q_2$, and to first order in $Q_2$, there are three conserved quantities: energy per unit mass $E= v^2/2 - U$, angular momentum per unit mass along the symmetry axis $\bm e$, $h_z = {\bm h} \cdot {\bm e}$ where $h = {\bm x} \times {\bm v}$, and a Carter-like constant
\begin{equation}
C =  h^2 +  \frac{Q_2}{m} \left [ ({\bm e} \cdot {\bm v})^2 - \frac{2Gm}{r} ({\bm e} \cdot {\bm n})^2 \right ] \,,
\end{equation}
where ${\bm n} = {\bm x}/r$.  (Since we are in Newtonian gravity, the conservation of rest mass $\mu$, which links coordinate time and proper time, is not relevant.)  They went on to show that, for motion in a field with any individual non-zero harmonic $Q_\ell$, with $\ell > 2$, no such constant exists.

However, this raises the question, in Newtonian gravitation, and for that matter in Maxwell electrostatics, does a Carter constant exist for   {\em general} axisymmetric distributions of mass (or charge)?  In this paper we point out that the answer is yes, if the moments satisfy the constraints 
\begin{subequations}
\begin{eqnarray}
Q_{2\ell +1} &=& 0  \,, \\
Q_{2\ell} &=& m \left (Q_2/m \right )^\ell  \,.
\end{eqnarray}
\label{condition}
\end{subequations}
Strangely, these are precisely the relations among the electric moments imposed by the no-hair theorem (\ref{nohair}) for the Kerr geometry.   

In this paper we prove this unusual property of Newtonian gravity (Sec. \ref{sec:Newt}) .  
If one augments Newtonian gravity with a gravitomagnetic field (or, in electrodynamics,  if one adds a multipolar magnetic field), we are not able to find an analogous Carter constant, unless the electric moments satisfy the conditions above, and all the gravitomagnetic moments (or magnetic moments) vanish (Sec. \ref{sec:mag}).  In Sec. \ref{sec:conclusion} we discuss the implications of this result.

\section{Carter-like constant of the motion in Newtonian gravity}
\label{sec:Newt}

The Newtonian equation of motion of a test body orbiting a massive body with a density distribution $\rho({\bm x})$ is
\begin{equation}
{\bm a} = d^2 {\bm x}/dt^2 = {\bm \nabla} U \,,
\end{equation}
where 
\begin{equation}
U = 
G\int \frac{\rho({\bm x}')}{|{\bm x} - {\bm x}'|} d^3x' \,,
\end{equation}
where $G$ is the Newtonian gravitational constant and $\rho({\bm x})$ is the mass density.  After suitable replacements involving charges, this is also the non-relativistic equation of motion of a charge in the electrostatic field of a charge distribution.

We Taylor expand $1/|{\bm x} - {\bm x}'|$ in powers of $x'$ and define
a set of symmetric, trace-free (STF) multipole moments $I^\stf{L}$, defined by
\begin{equation}
I^\stf{L} := \int \rho ({\bm x}') x^{'\stf{L}} d^3x' \,,
\end{equation}
where $L$ denotes an $\ell$-dimensional multi-index $ijk\dots$, and  ${x'}^{\stf{L}}$ denotes a product of $\ell$ components of $x'$, ${x'}^i {x'}^j \dots$, made trace-free.  For example ${x'}^{\stf{2}} = {x'}^i {x'}^j - {r'}^2 \delta^{ij}/3$.   See \cite{thorneRMP} for discussion and definitions of multipole fields and STF tensors.  We then obtain the equation of motion
\begin{eqnarray}
a^i 
&=&
-G\frac{mx^i}{r^3} + G\sum_{\ell =2}^\infty \frac{(-1)^\ell}{\ell !} I^\stf{L} \partial^{iL} \left ( \frac{1}{r} \right ) \,,
\label{eom}
\end{eqnarray}
where  $r = |{\bm x}|$,  and 
$\partial^{iL}$ denotes $\ell +1$ products of partial derivatives, with contraction performed over the $\ell$ repeated indices.  The mass and center of mass of the body are chosen such that $m = I^0 = \int \rho d^3x$, and $I^i = 0 = \int \rho x^i d^3x$.  

We assume that the central body is axisymmetric about an axis characterized by a unit vector $\bm e$.  It then follows that $I^\stf{L}$ for $\ell \ge 2$ must be proportional to an STF product of $\ell$ components of $\bm e$, in other words,
\begin{equation}
I^\stf{L} := Q_\ell  e^\stf{L} \,.
\label{ILdefined}
\end{equation}
The quantity $Q_\ell$ defines the mass or electric $\ell$-pole moment, and can be shown to be given by
\begin{equation}
Q_\ell = \int \rho({\bm x}) r^\ell P_\ell (z) d^3x \,,
\end{equation}
where $P_\ell (z)$ is a Legendre polynomial, and $z:={\bm e} \cdot {\bm n}$.

We make use of the well-known fact that
\begin{equation}
\partial^{iL} \left ( \frac{1}{r} \right ) = (-1)^{\ell +1} (2\ell +1)!! \frac{n^\stf{iL}}{r^{\ell+2}} \,.
\label{partial1overr}
\end{equation}
Then, substituting Eqs. (\ref{ILdefined})
and (\ref{partial1overr}) into the equation of motion (\ref{eom}), and exploiting the identity
\begin{equation}
e^\stf{L} n^\stf{iL} = \frac{\ell !}{(2\ell +1)!!} \left [ n^i P_{\ell +1}' (z) - e^i P_\ell' (z) \right ] \,,
\end{equation}
where prime denotes $d/dz$, we obtain the equation of motion
\begin{equation}
{\bm a} = -G\frac{m{\bm n}}{r^2} - \sum_{\ell =2}^\infty \frac{GQ_\ell}{r^{\ell+2}} \left [{\bm n} P_{\ell +1}' (z) - {\bm e} P_\ell' (z) \right ] \,.
\label{eom2}
\end{equation}

Contracting this equation with $\bm v$, using the fact that 
\begin{equation}
{\bm v} = \dot{r} {\bm n} + r \dot{\bm n} \,, \, dP_\ell/dt = (\dot{\bm n} \cdot {\bm e}) P'_\ell \,,
\label{vdef}
\end{equation} 
where an overdot denotes $d/dt$,
and using the recursion relation $P_{\ell +1}' = zP_\ell' + (\ell +1)P_\ell$, it is straightforward to find a constant of the motion
\begin{eqnarray}
\tilde{E} = \frac{1}{2} v^2 - \frac{Gm}{r} - \sum_{\ell =2}^\infty \frac{GQ_\ell}{r^{\ell+1}} P_\ell (z) 
= \frac{1}{2} v^2 - U({\bm x}) 
\,.
\end{eqnarray}
This of course is the energy per unit mass, conserved because the potential is stationary.
Defining the angular momentum per unit mass ${\bm h} := {\bm x} \times {\bm v}$, we find that 
\begin{equation}
\frac{d{\bm h}}{dt} =( {\bm x} \times {\bm e})\sum_{\ell =2}^\infty \frac{GQ_\ell}{r^{\ell+2}} P_\ell' (z) \,.
\end{equation}
Thus the component ${\bm h} \cdot {\bm e}$ of the angular momentum along the body's symmetry axis is conserved, a consequence of axisymmetry.  

But are there other non-trivial conserved quantities?   Given that the only vectors in the problem are $\bm x$, $\bm v$, $\bm h$ and $\bm e$, the only interesting possibility is the norm $h^2$, which itself is conserved only for spherical symmetry or for an orbit confined to the equatorial plane.  Calculating its rate of change, we find
\begin{eqnarray}
\frac{1}{2} \frac{d}{dt} h^2 &=& {\bm h} \cdot \frac{d{\bm h}}{dt}
\nonumber \\
&=& {\bm h} \cdot ( {\bm x} \times {\bm e}) \sum_{\ell =2}^\infty \frac{GQ_\ell}{r^{\ell+2}} P_\ell' (z)
\nonumber \\
&=& (\dot{\bm n} \cdot {\bm e}) \sum_{\ell =2}^\infty \frac{GQ_\ell}{r^{\ell-1}} P_\ell' (z) \,.
\label{hdot1}
\end{eqnarray}
Pulling out the $\ell =2$ term, inserting the relation (\ref{vdef}) between $\dot{\bm n}$ and $\bm v$, it can be shown that the $\ell =2$ term can be written as
\begin{eqnarray}
&&\frac{3GQ_2}{r^2} ({\bm n} \cdot {\bm e})\left [({\bm v} \cdot {\bm e}) - \dot{r} ({\bm n} \cdot {\bm e}) \right ] 
\nonumber \\
&=& \frac{d}{dt} \left [ \frac{GQ_2}{r} ({\bm n} \cdot {\bm e})^2 \right ]
+  \frac{Q_2}{m} \frac{Gm}{r^2} ({\bm n} \cdot {\bm e})({\bm v} \cdot {\bm e}) 
\nonumber \\
&=& \frac{d}{dt} \left [  \frac{Q_2}{m} \left ( \frac{Gm}{r} ({\bm n} \cdot {\bm e})^2 - \frac{1}{2} ({\bm v} \cdot {\bm e})^2 \right ) \right ]
\nonumber \\
&& -  \frac{Q_2}{m}({\bm v} \cdot {\bm e})  \sum_{\ell =2}^\infty \frac{GQ_\ell}{r^{\ell+2}} (\ell +1)P_{\ell+1} (z) \,,
\end{eqnarray}
where we used the equation of motion (\ref{eom2}) to eliminate $Gm{\bm n}/r^2$, pulled out other time derivatives, and used the recursion relation $P_\ell' = zP_{\ell+1}' -(\ell +1)P_{\ell +1}$.  Combining this with the remaining terms in the sum in Eq.\ (\ref{hdot1}) and again using Eq.\ (\ref{vdef}), as well as further identities satisfied by the Legendre polynomials, we obtain finally,
\begin{eqnarray}
 &&\frac{d}{dt}
 \left [ \frac{1}{2}h^2 + \frac{1}{2}\frac{Q_2}{m} ({\bm v} \cdot {\bm e})^2 - \frac{Q_2}{m}({\bm n} \cdot {\bm e}) \sum_{\ell=0}^\infty  \frac{GQ_\ell}{r^{\ell+1}} P_{\ell+1} (z) \right .
 \nonumber \\
  &&\left .\quad \quad-
   \sum_{\ell=2}^\infty \frac{GP_\ell}{r^{\ell -1}} \left ( Q_\ell - \frac{Q_2}{m} Q_{\ell -2} \right ) \right ]
 \nonumber \\
 &&  \qquad =\dot{r} \sum_{\ell=2}^\infty (\ell -1) \frac{GP_\ell}{r^{\ell}} \left ( Q_\ell - \frac{Q_2}{m} Q_{\ell -2} \right )  \,.
 \label{finalconstant1}
\end{eqnarray}

No amount of further manipulation can convert the expression on the right hand side into a total time derivative.  Since the Legendre polynomials are linearly independent, then for arbitrary orbits, there will be a constant of the motion of the chosen form if and only if 
\begin{equation}
Q_\ell = \frac{Q_2}{m} Q_{\ell -2}\,, \,  {\rm for \, all} \, \ell \ge 2 \,.
\end{equation}
Since $Q_1 =0$, this implies Eq.\ (\ref{condition}).  The Carter constant of the motion is then given by
\begin{equation}
C = h^2 + \frac{Q_2}{m} ({\bm v} \cdot {\bm e})^2 - 2{\bm n} \cdot {\bm e} \sum_{\ell=0}^\infty  \frac{m(Q_2/m)^{\ell+1} }{r^{2\ell+1}} P_{2\ell+1} (z) \,.
\end{equation}
This generalizes Eq. (2.8) of \cite{flanhind}.

What kind of Newtonian source satisfies the conditions of Eq.\ (\ref{condition})?  Its potential takes the form
\begin{eqnarray}
U({\bm x}) &=& \frac{Gm}{r} + \sum_{\ell =1}^\infty \frac{GQ_{2\ell}}{r^{2\ell +1}} P_{2\ell} (z) 
\nonumber \\
&=& \frac{Gm}{r} \sum_{\ell =0}^\infty \left (\frac{Q_2/m}{r^2} \right )^\ell P_{2\ell} (z) \,.
\end{eqnarray}
Using the generating function for Legendre polynomials, $\sum_{\ell =0}^\infty t^\ell P_\ell (z)= (1-2tz+z^2)^{-1/2}$, we can rewrite $U$ in the form
\begin{equation}
U({\bm x}) = \frac{Gm/2}{|{\bm x} - \alpha {\bm e} |}
+\frac{Gm/2}{|{\bm x} + \alpha {\bm e} |} \,,
\end{equation}
where $\alpha := (Q_2/m)^{1/2}$.  If $Q_2/m$ is positive, which corresponds to the prolate case, then this is the potential of two point sources, each of mass $m/2$, separated by a distance $2\alpha$ along the $\bm e$ axis.  In the oblate case, $\alpha$ is imaginary, so the potential, while real, does not have an interpretation in terms of simple point masses.

\section{Newtonian gravity plus gravitomagnetism}
\label{sec:mag}

We now add gravitomagnetism to Newtonian theory.  This is a model of gravity that encapsulates some, though not all, of the features of linearized general relativity (see \cite{bct}, for example).  In electromagnetism, it is the non-relativistic limit of the Lorentz-Dirac equation. The equation of motion in this case takes the form
\begin{equation}
{\bm a}  = {\bm \nabla} U + \frac{1}{c}{\bm v} \times \left ( {\bm \nabla} \times {\bm A}_g \right ) \,,
\end{equation}
where ${\bm A}_g$ is the gravitomagnetic potential, given by
\begin{eqnarray}
{\bm A}_g^i &=& -\frac{4G}{c} \int \frac{\rho ({\bm x}') v'^i }{|{\bm x} - {\bm x}'|} d^3x' \,,
\nonumber \\
&=&-\frac{4G}{c} \sum_{\ell =0}^\infty \frac{(-1)^\ell (\ell+1)}{(\ell+2)!} \epsilon^{iqp} {J}^{\stf{pL}} \partial^{qL} \left ( \frac{1}{r} \right ) ,
\end{eqnarray}
where ${J}^{\stf{pL}}$ are STF current multipole moments, given, for a stationary axisymmetric body by $J_{\ell +1} e^\stf{pL}$; here $J_1=S$ is the ordinary angular momentum (or the magnetic moment) of the body. 

This generates a term to be added to the equation of motion (\ref{eom}) given by
\begin{equation}
\delta {\bm a} = \frac{4G}{c^2} \sum_{\ell=1}^\infty \frac{\ell}{\ell+1} \frac{J_\ell}{r^{\ell +2}} \left [{\bm n} P_{\ell +1}' (z) - {\bm e} P_\ell' (z) \right ] \times {\bm v} \,.
\end{equation}
This gravitomagnetic term does not affect the energy, but it does generate a new conserved angular momentum given by
\begin{equation}
L_z = {\bm h} \cdot {\bm e} - \frac{4G}{c^2} \sum_{\ell=1}^\infty
\frac{J_\ell [1-({\bm e} \cdot {\bm n})^2]}{(\ell +1)r^\ell} P'_\ell (z) \,.
\end{equation}

However, in this case, we find no Carter-like constant in general.   Our attempt at construction included using a linear combination of $h^2$, $L_z^2$ and $L_z\, S$.  We find that a Carter-like constant exists if conditions (\ref{condition}) are met, along with $J_\ell = 0$ for all $\ell$, which is our original purely Newtonian case.  If $S \ne 0$, and if we work only to first order in multipole moments, then we find a Carter constant given by
\begin{equation}
C= h^2 + \frac{Q_2}{m} ({\bm v} \cdot {\bm e})^2 - 2\frac{Q_2}{r}({\bm n} \cdot {\bm e})^2 - \frac{4S}{cr} {\bm h} \cdot {\bm e} \,,
\end{equation}
which is consistent with Eq. (2.23) of \cite{flanhind}.
The absence of a constant in the general gravitomagnetic case highlights the very special nature of geodesic motion in the Kerr geometry.

\section{Discussion}
\label{sec:conclusion}

In a purely Newtonian context, this result may not have much practical significance, given the contrived nature of the source that possesses a Carter constant.  However, it does raise a number of related questions.

The first is purely mathematical: in general relativity, there is an analogue to the solution for the field of two point masses separated by a fixed distance.  Known as the Bach-Weyl solution~\cite{bachweyl,skmhh}, it is an exact, static axisymmetric vacuum solution of Einstein's equation, with the property that it is singular on a line joining the two masses; the singularity represents a strut required to hold the two masses apart.   Does the Bach-Weyl solution possess a Carter-like constant, or equivalently, does it admit an appropriate non-trivial Killing tensor?

In Newtonian gravity, does there exist a physically reasonable, oblate distribution of matter (with $Q_2 < 0$) that satisfies Eq.\ (\ref{condition})?

What happens when the equations are extended to higher post-Newtonian orders?  The gravitomagnetic model considered above included only terms linear in $S = ma$.  If corrections to the equations of motion of order $S^2$ are included, can a Carter constant be found at some PN order and at higher order in the moments~\cite{bltw}?  Considerations such as these may elucidate what it is about Kerr that makes a Carter constant possible, and may also give insights into how to parametrize spacetimes that are ``not quite'' Kerr, in order to see how observations of EMRIs could be used to test general relativity in the strong-field regime~\cite{hughes}.

\begin{acknowledgments}

This work was supported in part by the National Science Foundation, Grant No.\ PHY 06--52448, the National Aeronautics and Space Administration, Grant No.\
NNG-06GI60G, and the Centre National de la Recherche Scientifique, Programme Internationale de la Coop\'eration Scientifique (CNRS-PICS), Grant No. 4396.
We are grateful for the hospitality of the Institut d'Astrophysique de Paris, where this work was carried out.  We thank Emanuele Berti, Luc Blanchet, \'Eanna Flanagan, Malcolm MacCallum, Alexandre Le Tiec, and Adamantios Stavridis for helpful comments.
\end{acknowledgments}

\vfill


\begin{thebibliography}{}

\bibitem{chandra}
S. Chandrasekhar,
{\em The Mathematical Theory of Black Holes} (Clarendon Press, Oxford, 1983).

\bibitem{mtw}
C. W. Misner, K. S. Thorne and J. A. Wheeler,
{\em Gravitation}
(Freeman, San Francisco, 1973).

\bibitem{carter72}
B. Carter, Phys.\ Rev.\ Lett.\  {\bf 26}, 331(1972).

\bibitem{robinson}
D. C. Robinson, Phys.\ Rev.\ Lett.\  {\bf 34}, 905 (1975).

\bibitem{hansen}
R. O. Hansen, J.\ Math.\ Phys.\ {\bf 15}, 46 (1974).

\bibitem{carter}
B. Carter, Phys.\ Rev.\ {\bf 174}, 1559 (1968).

\bibitem{defelice}
F. de Felice and G. Preti, Class.\ Quantum Gravit.\ {\bf 16}, 2929 (1999).

\bibitem{rosquist}
K. Rosquist, T. Bylund and L. Samuelsson, preprint (arXiv: 0710.4260).

\bibitem{walker}
M. Walker and R. Penrose, Commun.\ Math.\ Phys.\ {\bf 18}, 265 (1970).

\bibitem{mino}
Y. Mino, Phys.\ Rev.\ D {\bf 67}, 084027 (2003).

\bibitem{drasco05}
S. Drasco, \'E. \'E. Flanagan and S. A. Hughes, Class.\ Quantum Gravit.\ {\bf 22}, S801 (2005).

\bibitem{hughes05}
S. A. Hughes, S. Drasco, \'E. \'E. Flanagan and J. Franklin, Phys.\ Rev.\ Lett.\ {\bf 94}, 221101 (2005).

\bibitem{sago}
N. Sago, T. Tanaka, W. Hikida and H. Nakano, Prog.\ Theor.\ Phys.\ {\bf 114} 509 (2005).

\bibitem{flanhind}
\'E. \'E. Flanagan and T. Hinderer,
Phys.\ Rev.\ D {\bf 75}, 124007 (2007).

\bibitem{thorneRMP}
K. S. Thorne, Rev.\ Mod.\ Phys.\ {\bf 52},  300 (1980).

\bibitem{bct}
V. B. Braginsky, C. M. Caves and K. S. Thorne, Phys.\ Rev.\ D {\bf 15}, 2047 (1977).

\bibitem{bachweyl}
R. Bach and H. Weyl, Math.\ Z.\ {\bf 13}, 134 (1922).

\bibitem{skmhh}
H. Stephani, D. Kramer, M. MacCallum, C. Hoenselaers and E. Herlt,
{\em Exact Solutions of Einstein's Field Equations}, (Cambridge University Press, Cambridge, 2003), pp 304 - 307.

\bibitem{bltw}
L. Blanchet, A. Le Tiec and C. M. Will, work in progress.

\bibitem{hughes}
S. A. Hughes, in
{\em Laser Interferometer Space Antenna: 6th International LISA Symposium}, AIP Conference Proceedings, Vol 873,  ed. S. M. Merkowitz and J. C. Livas (American Institute of Physics, Melville, NY 2006), p. 233.

\end{thebibliography}
\end{document}